\documentclass[iop]{emulateapj}

\usepackage{graphicx}
\usepackage{textcomp}
\usepackage{natbib}

\def\hi{H\,\textsc{i}}
\def\nh{$N_{\rm{H{\scriptscriptstyle\,I}}}$}

\begin{document}

\title{First Connection between Cold Gas in Emission and Absorption: CO Emission from a Galaxy-Quasar Pair}

\shorttitle{CO emission from an Galaxy-Quasar Pair}
\shortauthors{Neeleman et al.}

\author{Marcel Neeleman\altaffilmark{1}, J. Xavier Prochaska\altaffilmark{1}, Martin A. Zwaan\altaffilmark{2},
Nissim Kanekar\altaffilmark{3}, Lise Christensen\altaffilmark{4}, Miroslava Dessauges-Zavadsky\altaffilmark{5}, 
Johan P.U. Fynbo\altaffilmark{4}, Eelco van Kampen\altaffilmark{2}, Palle M{\o}ller\altaffilmark{2}, 
Tayyaba Zafar\altaffilmark{6}}
\altaffiltext{1}{Department of Astronomy \& Astrophysics, UCO/Lick Observatory, University of California, 
1156 High Street, Santa Cruz, CA 95064, USA; marcel@ucsc.edu}
\altaffiltext{2}{European Southern Observatory, Karl-Schwarzschild-strasse 2, D-85748 Garching bei M\"{u}nchen, Germany}
\altaffiltext{3}{Swarnajayanti Fellow; National Centre for Radio Astrophysics, Tata Institute of Fundamental Research, Pune 411007, India}
\altaffiltext{4}{Dark Cosmology Centre, Niels Bohr Institute, University of Copenhagen, Juliane Maries Vej 30, DK-2100 Copenhagen, Denmark}
\altaffiltext{5}{Observatoire de Gen\`{e}ve, Universit\'{e} de Gen\`{e}ve, 51 Ch. des Maillettes, 1290 Sauverny, Switzerland}
\altaffiltext{6}{Australian Astronomical Observatory, P.O. Box 915, North Ryde, NSW 1670, Australia}

\begin{abstract} 
We present the first detection of molecular emission from a galaxy selected to be near a projected
background quasar using the Atacama Large Millimeter/submillimeter Array (ALMA). The ALMA detection of 
CO(1$-$0) emission from the $z=0.101$ galaxy toward quasar PKS 0439-433 
is coincident with its stellar disk and yields a molecular gas 
mass of $M_{\rm mol} \approx 4.2 \times 10^9 M_\odot$ (for a Galactic CO-to-H$_2$ conversion factor), 
larger than the upper limit on its atomic gas mass. We resolve the CO velocity field, obtaining a 
rotational velocity of $134 \pm 11$\,km\,s$^{-1}$, and a resultant dynamical mass of 
$\geq 4 \times 10^{10}$\,M$_\odot$. Despite its high metallicity and large molecular mass, 
the $z=0.101$ galaxy has a low star formation rate, implying a large gas consumption timescale, 
 larger than that typical of late-type galaxies. Most of the molecular gas is hence likely to be in 
 a diffuse extended phase, rather than in dense molecular clouds. 
 By combining the results of emission and absorption studies, 
we find that the strongest molecular absorption component toward the quasar cannot arise from 
the molecular disk, but is likely to arise from diffuse gas in the galaxy's circumgalactic medium. 
Our results emphasize the potential of combining molecular and stellar emission line studies 
with optical absorption line studies to achieve a more complete picture of the gas within and 
surrounding high-redshift galaxies.
\end{abstract}


\section{Introduction}
\label{sec:intro}
A crucial link in our understanding of galaxy formation and evolution is the 
recognition of the pivotal role that gas plays in the process. In a $\Lambda$ cold dark matter
cosmology, initial small dark matter overdensities merge to become large overdensities
in which galaxies can grow. However, it is the baryonic 
component in and surrounding this nascent galaxy, in the form of gas,
that shapes and nurtures the galaxy, feeding its growth, and acting as the fuel 
for star formation. Studies of gas in and around galaxies are hence 
critical to understand galaxy evolution. 

Absorption studies have long provided the best route toward probing the gaseous environs of 
galaxies at high redshifts. Using samples of galaxies with a nearby projected background quasar 
(i.e. ``quasar$-$galaxy pairs''), the circumgalactic medium (CGM) of different galaxy populations can
be probed \citep[e.g.,][]{Hennawi2006,Stocke2013,Tumlinson2013,Bordoloi2014}. These studies 
find that the CGM extends out to at least 100\,kpc around the most massive galaxies 
\citep{Prochaska2011}, contains a large fraction of both cool and hot ionized gas \citep{Werk2014}, 
and is metal-enriched \citep[e.g.][]{Simcoe2006,Lau2016}. Conversely, selecting samples of 
strong absorbers through
direct quasar spectroscopy allows for the study of the gas in and surrounding more representative 
galaxies, as the usual biases in identifying galaxy samples are not employed here \citep[e.g.,][]{Wolfe2005}.

As part of an Atacama Large Millimeter/submillimeter Array (ALMA) program to study molecular gas in 
high-metallicity, absorption-selected galaxies, we targeted the strong metal-line absorber at 
$z = 0.101$ toward PKS0439$-$433 \citep{Petitjean1996}. This super-solar metallicity absorber is 
associated with a spiral galaxy, at a projected distance of 7.3\,kpc to the quasar 
\citep{Petitjean1996,Chen2005}, providing an ideal quasar$-$galaxy pair for studying the molecular 
gas in the galaxy, as well as its CGM. Besides providing a detailed characterization of the 
host galaxy, our ALMA observations indicate that the absorber is probing a molecular gas component 
of galaxies that is essentially impossible to observe in emission. Our study thus highlights the 
invaluable information that can be obtained by studying the galactic counterparts of absorbers 
in constraining how galaxies form and evolve. 

\section{The $z = 0.101$ absorber toward PKS 0439$-$433}
\label{sec:0439}

The $z = 0.101$ absorber toward PKS 0439$-$433, has been studied in detail in 
multiple wavebands, using both imaging and spectroscopy. The {\hi} column density of the absorber
has been measured from the Lyman-$\alpha$ absorption line to be 
log[\nh/cm$^{-2}$]~$=19.63 \pm 0.08$ \citep{Muzahid2015,Som2015}. The detection of several low-ionization 
metal lines, in particular S\,\textsc{ii}, indicates that the absorber has two main absorption complexes 
at $z_1=0.10094$ and $z_2=0.10119$, with  a super-solar gas-phase metallicity, [S/H]=$0.28\,\pm\,0.08$ 
\citep{Dutta2015}. These two main absorption components are also seen in Ca\,\textsc{ii} and 
Na\,\textsc{i} absorption toward the quasar \citep{Richter2011}.

The system furthermore shows ultraviolet H$_2$ absorption, with H$_2$ column density $\approx(4.1 \pm 0.5) 
\times 10^{16}$~cm$^{-2}$ \citep{Muzahid2015}, and also has a tentative 
(3.3$\sigma$) detection of {\hi}~21\,cm absorption, yielding a low spin temperature, $\approx 90$~K
\citep{Kanekar2001,Dutta2015}, although the {\hi}~21\,cm feature does not coincide in velocity 
with the metal lines.

\begin{figure}[t!]
\includegraphics[width=0.5\textwidth]{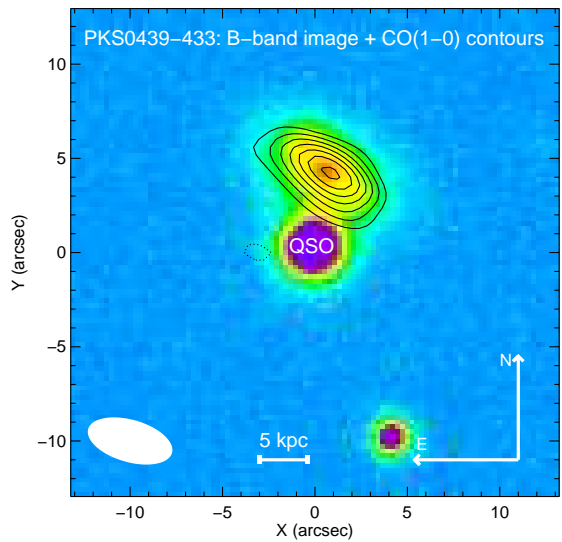}
\caption{Integrated CO(1$-$0) emission, in contours, overlaid on a B-band image of the field 
surrounding PKS~0439$-$433 taken by \citet{Chen2005}. The outermost CO contour is at 3$\sigma$ 
significance, with each successive contour increasing by $\sqrt 2$-$\sigma$. It is clear that the CO(1$-$0) 
emission arises from the disk of the $z=0.101$ spiral galaxy.}
\label{fig:img}
\end{figure}

Optical spectroscopy of the $z = 0.101$ spiral galaxy has identified multiple emission lines, 
including H$\alpha$, H$\beta$, [O\,{\textsc{ii}], etc. \citep{Petitjean1996,Chen2005}. 
\citet{Chen2005} used slit spectroscopy of the H$\beta$ line to find that the galaxy's velocity 
field is consistent with arising from a rotating disk at an inclination of $58$\textdegree $\pm~5$\textdegree.
The H$\alpha$ luminosity was used by \citet{Dutta2015} to estimate a star formation rate (SFR) 
of $0.53\,M_\odot$\,yr$^{-1}$ based on a Salpeter IMF, without including corrections due 
to slit-losses of the H$\alpha$ emission.

Finally, the stellar mass of the spiral galaxy has been estimated to be log[$M_\star/M_\odot$]\,
$=10.01 \pm 0.02$, via a fit of spectral energy distribution models to the 
optical photometry \citep{Christensen2014}. This also yielded an estimate for the SFR of
$\approx1.5\,M_\odot$\,yr$^{-1}$ without correcting for intrinsic extinction. The non-detection 
of {\hi}~21\,cm emission from the galaxy yields a 3-$\sigma$ upper limit of $3 \times 10^9\,M_\odot$ 
on its atomic gas mass \citep{Kanekar2001}.
 
\section{ALMA observations}
\label{sec:obs}

We used ALMA band~3 receivers to carry out a deep search for redshifted CO(1$-$0) emission
from the $z = 0.101$ galaxy on UT~2014~December~25. A compact antenna configuration resulted in an 
angular resolution of $\sim 2\,''$, corresponding to $\approx 3.7$\,kpc at $z=0.101$. One of the four 
spectral windows was centered at the redshifted CO(1$-$0) line frequency of 
104.7\,GHz, while the three other spectral windows were used for continuum 
observations of the field. QSO~J0439$-$4522 and Uranus were used as gain and amplitude calibrators, respectively.

The data were analyzed following standard procedures in the Common Astronomy Software Applications 
package \citep[CASA;][]{McMullin2007}. The continuum image was made by combining the three 
continuum spectral 
windows, resulting in a root mean square (RMS) noise of 61\,$\mu$Jy. The quasar was detected at high 
signal-to-noise ratio in this image, allowing for self-calibration of the visibility data by performing one round
of phase only and one round of phase and amplitude calibration. After self-calibration, the final 
spectral cube covering the CO(1$-$0) line was made with a velocity resolution
of 20\,km\,s$^{-1}$, using Briggs weighting with a robust factor of 0.5 and the {\tt clean} routine. 
The synthesized beam size of the cube was $4.28\,'' \times 1.96\,''$, at a position angle of 73.85\textdegree. 
The final RMS noise of the data cube is $0.76$\,mJy\,beam$^{-1}$ per 20\,km\,s$^{-1}$ channel.

\begin{figure}[b!]
\includegraphics[width=0.5\textwidth]{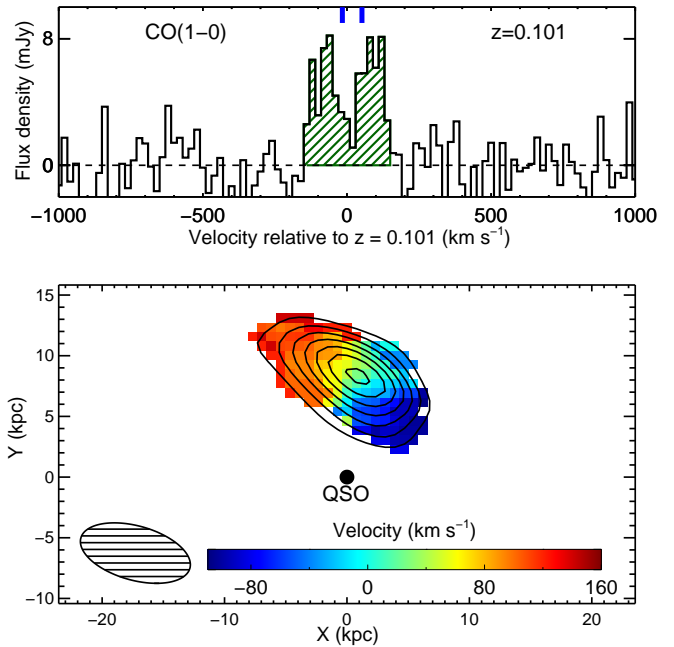}
\caption{Top:~integrated flux density as a function of the velocity, where $v =0$\,km\,s$^{-1}$ corresponds
to $z=0.10100$. Blue tick marks indicate the velocity centroid of the two absorption components.
Bottom:~velocity field of the CO(1$-$0) emission from the $z=0.101$ spiral galaxy. 
See the text for a discussion.}
\label{fig:vel}
\end{figure}

\begin{figure*}[t!]
\includegraphics[width=\textwidth]{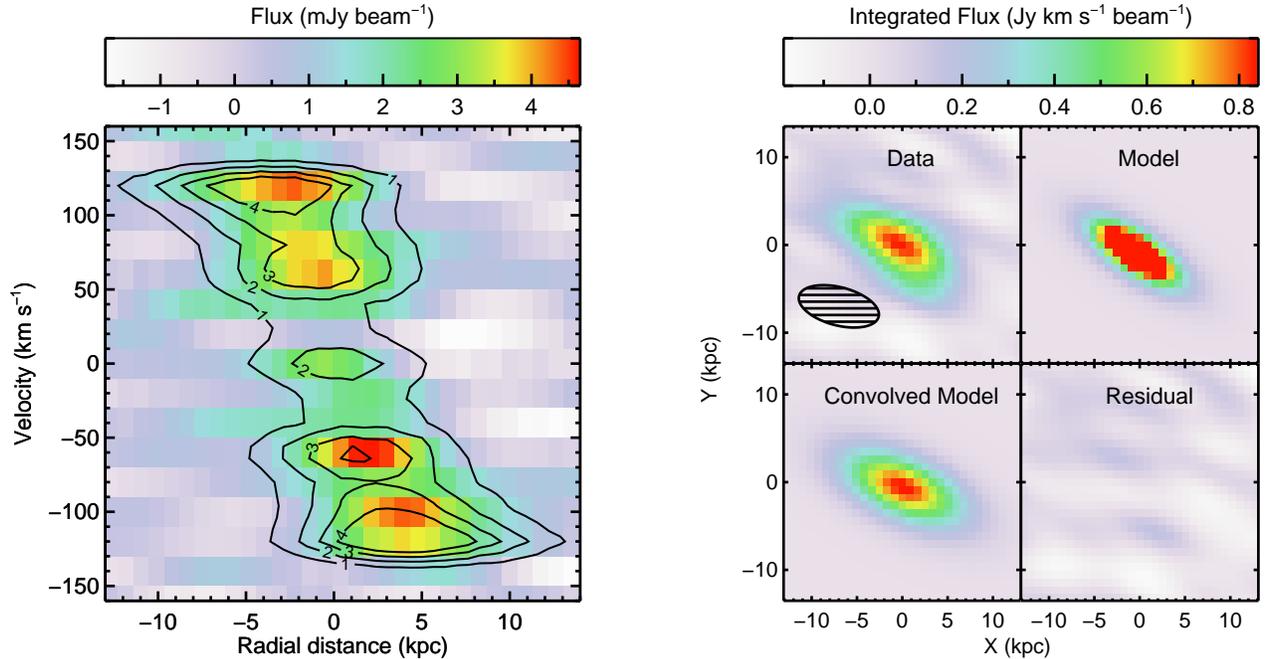}
\caption{Left:~position-velocity ($p-v$) diagram of the CO(1$-$0) line, for a cut along the
major axis of the galaxy. Overplotted on the data, in contours, is the best-fit 
model from the Monte Carlo Markov Chain method described in the text convolved with the ALMA 
beam. Right:~comparison of the integrated CO(1$-$0) line emission for the data, model, and 
convolved model. The residual and $p-v$ diagrams show that the simple infinitely thin disk-model 
describes the data well.}
\label{fig:pv}
\end{figure*}
 
\section{Results}
\label{sec:results}

A search for line emission in the spectral cube yields a strong (12$\sigma$) detection of 
emission, at a frequency corresponding to $z=0.10100$ for the CO(1$-$0) line. Figure~\ref{fig:img} shows 
the integrated CO(1$-$0) emission overplotted on the B-band optical image of \citet{Chen2005}. 
The CO emission is centered on the spiral galaxy previously associated with the 
absorber \citep{Petitjean1996,Chen2003}, with excellent agreement in both the alignment and the 
spatial extent of the CO and the optical emission. We obtain an integrated CO(1$-$0) flux density of 
$1.46 \pm 0.07$\,Jy\,km\,s$^{-1}$, which results in a line luminosity of $L'_{\rm{CO}} = 7.0 \pm 0.3
\times 10^8$\,K\,km\,s$^{-1}$\,pc$^2$, or $L_{\rm{CO}} = 3.4 \pm 0.2 \times 10^{4}\,L_\odot$.

No CO(1$-$0) absorption was detected toward the quasar, yielding a $3\sigma$ upper limit of
$\approx 0.027$~km\,s$^{-1}$ on the integrated line opacity. Assuming a line excitation temperature 
of $\approx 10$~K and a Galactic CO-to-H$_2$ conversion factor typical of diffuse/translucent clouds 
\citep[N(CO)/N(H$_2$) $\approx 3 \times 10^{-6}$;][]{Burgh2007}, we obtain 
N(H$_2$)~$< 6 \times 10^{19}$~cm$^{-2}$, consistent with the estimated H$_2$ column
density from the ultraviolet lines.

The integrated CO(1$-$0) emission profile is displayed in the top panel of Figure~\ref{fig:vel}
Furthermore, the emission is spatially resolved even in ALMA's compact configuration, allowing 
an estimate of the velocity field of the galaxy. The bottom panel of Figure~\ref{fig:vel} shows the 
first velocity moment of the CO(1$-$0) emission obtained by using the {\tt immoments} function 
in CASA, and by only including emission detected at greater than 3$\sigma$ significance. 
Both the `spider' pattern in this figure and the `double-horned' feature in the integrated line profile
indicate the CO(1$-$0) emission arises from a rotating disk \citep[e.g.,][]{DeBlok2014}, strongly suggesting that 
the emission originates from the interstellar medium of the spiral galaxy. 

To quantitatively describe the results, one must take into account several observational effects, 
most importantly beam smearing and inclination. To disentangle these effects, we have developed 
a custom Monte Carlo Markov Chain method, which we describe in the next section.
 
 \subsection{Modeling the CO(1$-$0) Emission}
 
 We model the CO(1$-$0) emission from the $z=0.101$ galaxy by an infinitely thin, exponential disk. 
 For simplicity, we assume a flat rotation curve, such that the rotational velocity of the molecular gas 
 does not depend on radius. We note that the intrinsic velocity dispersion for this model is zero, 
 the implications of which we will discuss later.
 
 To account for beam smearing, we adopt the following procedure. First, we construct a model 
 data cube by modeling the average CO emission as an exponential disk of the form
 \begin{equation} 
 I({\rm Jy\,beam}^{-1})=I_0\,e^{(-R/R_D)} \:\:,
 \end{equation}
where $I_0$ is the intensity of the emission line at the galaxy's center and $R_D$ is the turnover 
radius. The disk's orientation is determined by its inclination ($i$) and the position angle of 
the major axis with respect to north ($\alpha$). The galaxy's center in the data cube is 
described by three coordinates, two spatial, and one frequency. Finally, the rotation of 
the molecular disk is modeled by a constant rotational velocity along the disk ($v_{\rm max})$. 

The resultant model data cube is then convolved with the ALMA beam and compared
with the data set. We apply a Monte Carlo Markov Chain method using the Metropolis$-$Hastings algorithm to 
determine the posterior probability distribution function of each parameter of the model. During this process, 
we remove the `burn-in' phase by discarding the first 40\,\% of the chain. We check for convergence by
choosing five different starting points and running five different iterations.

From the residuals in Figure~\ref{fig:pv}, we can see that this simple model describes the data quite 
well. The residuals show no evidence of sub-structures, such as warped disks, although higher-
resolution data would be required to confirm this. For the model, we find an inclination 
angle of 67$^{+6}_{-5}$ degrees, with a position angle of $56$\textdegree$~\pm~5$\textdegree, and a rotational 
velocity of 134$^{+8}_{-11}$\,km\,s$^{-1}$. These values are in good agreement with those derived
from the H$\beta$ line \citep{Chen2005}.

\begin{figure}[b!]
\includegraphics[width=0.5\textwidth]{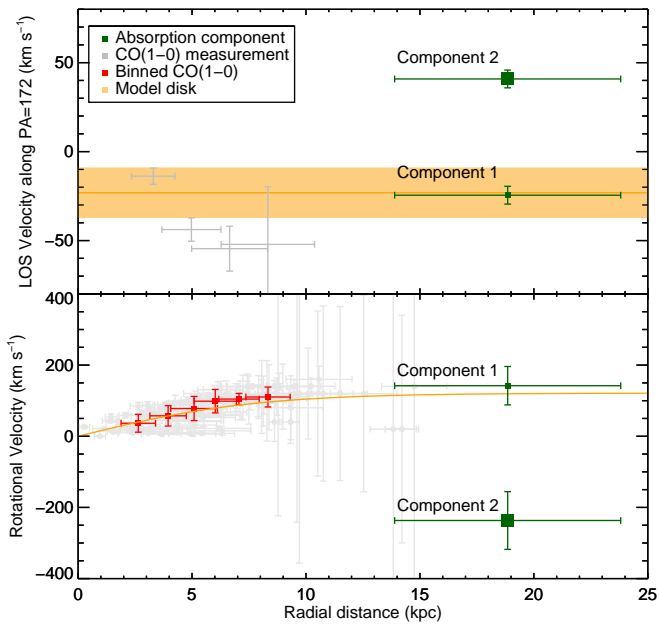}
\caption{Top:~line of sight (LOS) velocity component along PA = 172\.\textdegree. The gray 
points are measurements of the LOS velocity of the CO emission, whereas the orange shaded 
region is the LOS velocity along this PA for the best-fit model. The green data points are the two
absorption components. Bottom:~rotation curve of the molecular disk. The CO(1$-$0) emission 
profile (with raw and binned measurements in gray and 
red, respectively) is well described by a model with constant rotational velocity (orange). Both plots show that the 
main absorption complex (component~2) is inconsistent with coming from an extended molecular disk.}
\label{fig:disk}
\end{figure}
 
\subsection{Kinematic Structure of the Molecular Gas}

As shown previously, the CO(1$-$0) emission profile is consistent with arising
from a rotating molecular disk. To assess if the molecular gas seen in absorption
also arises from this disk, we can calculate the rotational velocity needed to produce
the velocity shifts seen in the absorption components. This is shown in Figure~\ref{fig:disk}.
It is clear that the velocity of absorption component~1 is consistent with arising from 
the disk. However, the dominant molecular gas resides in component~2, whose velocity 
is inconsistent with arising from an extended molecular disk.

Therefore, we conclude that absorption component~2 is unlikely to probe the same 
molecular disk seen in emission and must instead be probing molecular gas in either 
the CGM, a strong metal-rich outflow, or a nearby faint dwarf galaxy. The high metallicity of the gas 
favors the first two scenarios as a clear correlation between metallicity and mass 
exists, both for galaxies \citep[e.g.,][]{Tremonti2004} and absorbers 
\citep{Ledoux2006,Prochaska2008,Kanekar2009,Moller2013,Neeleman2013,Christensen2014}.
The velocity measured from the gas results in very modest outflow speeds 
($\lesssim 25$\,km\,s$^{-1}$), blurring the distinction between outflow and CGM. We 
therefore assert that the dominant H$_2$ absorption component stems from 
the CGM of the galaxy, in agreement with the results of \citet{Dutta2015}, based on
photoionization modeling of H$_2$ and Na\,\textsc{i} of the absorber.

\subsection{The Gas and Dynamical Mass of the Galaxy}

The measured CO(1$-$0) line luminosity can be converted to a molecular gas mass by 
assuming a CO-to-H$_2$ conversion factor, $\alpha_{\rm CO}$, and including a correction 
for helium. For this purpose, we will assume $\alpha_{\rm CO} = 4.3\,M_\odot$
\,(K\,km\,s$^{-1}$\,pc$^2$)$^{-1}$, as recommended by \citet{Bolatto2013} for normal galaxies 
of solar metallicity. This implies a total molecular gas mass of $M_{\rm mol} = 4.2 \pm 0.2 \times 
10^9\, M_\odot$, which is higher than the 3$\sigma$ upper limit on the 
neutral atomic gas mass \citep[$\leq 3 \times 10^9 M_\odot$;][]{Kanekar2001}. The mass ratio of 
molecular to atomic gas is $\geq 1.3$, placing this galaxy at the top end of the distribution 
compared to other late-type galaxies of comparable stellar mass
\citep[e.g.,][]{Lisenfeld2011,Saintonge2011,Boselli2014,Bothwell2014,Jiang2015}. 

The molecular gas fraction relative to stars ($M_{\rm mol}/M_\star$) is also quite high, $\approx 0.4$; 
for comparison, \citet{Boselli2014} obtain a typical molecular gas fraction of $\approx 0.1$ 
for late-type galaxies at low redshifts, although again this fraction falls within the expected
spread of late-type galaxies. Combining the total molecular gas mass with the estimated SFR of 
$1.5\,M_\odot$\,yr$^{-1}$ yields a gas consumption timescale of $\approx 2.8$\,Gyr, larger 
than the typical gas consumption timescales ($\approx 1$\,Gyr) seen in late-type galaxies 
of similar stellar mass \citep[e.g.,][]{Boselli2014}. The low SFR suggests that the molecular 
gas in the galaxy may be dominated by diffuse gas rather than by dense giant molecular clouds. 
Observations of dense gas tracers such as HCN or HCO$^+$ would be of much interest in this regard.

One can also use the rotation velocity to infer the dynamical mass of the galaxy, for an 
assumed dark matter density profile. Ignoring the details of the density profile, the dynamical 
mass can be roughly estimated as $M_{\rm dyn} \sim V^2 R/G$, where $V$ is the rotation velocity
and $R$ is the radius of the gas disk. In the present case, we have $V \approx 134$~km~s$^{-1}$ 
and $R \approx 10$\,kpc, yielding $M_{\rm dyn} \gtrsim 4 \times 10^{10}\,M_\odot$. This 
is likely to be a lower limit to the dynamical mass as the radius $R$ has been estimated from 
the molecular disk (whose size is similar to that of the stellar disk), which is likely
significantly less extended than the atomic gas \citep[e.g.,][]{Cayatte1994}.

\subsection{Surface Mass Density of the Molecular Gas}

Our exponential disk model yields an estimate of the CO intensity per unit area as a function of 
radius, which can be converted to a surface mass density with the assumption that the intensity scales 
with the mass of molecular gas, i.e. M$_{\rm H_2} =\alpha_{\rm CO} L'_{\rm CO}$. The molecular absorption 
yields an additional constraint on the surface mass density at the position where the line of sight 
crosses the disk plane. 

Figure~\ref{fig:surfh2} shows that the molecular gas seen in absorption does not fit a simple extrapolation
of the exponential disk model. The extrapolation of the model to where the sightline crosses the 
plane of the disk ($\approx 20$\,kpc) predicts an average surface density
several orders of magnitude higher than that measured in absorption. This indicates that the 
absorption components are tracing a more diffuse molecular gas than that responsible for the 
CO(1$-$0) emission.

We can estimate the total amount of molecular gas probed by the absorption component, if we make the following 
three assumptions: (i)~the covering fraction of this gas is small, $<10$\,\% \citep[this is corroborated 
by the paucity of H$_2$ absorption in absorption-selected galaxies; e.g.,][]{Ledoux2003, Noterdaeme2008}; 
(ii)~the measured H$_2$ column density toward PKS 0439$-$433 is representative of the whole 
distribution of gas; and (iii)~only gas within a sphere of 20\,kpc of the galaxy is considered. Under 
these assumptions, we find that the absorbing molecular component contains $\lesssim10^5\,M_\odot$ 
of gas. This is negligible in comparison to the mass of the galaxy and would be 
undetectable in emission at cosmological distances even with ALMA. 

The presence of molecular gas in the CGM indicates that a small fraction of the gas is dense and 
cold enough for molecular gas to be shielded from dissociating radiation. Such gas could be formed 
elsewhere and survived the transport into the CGM as is the favored explanation for H$_2$ gas seen 
locally in high-velocity clouds \citep[e.g.,][]{Sembach2001}, or could have formed in situ onto dust grains. 
A detailed analysis of these two scenarios is beyond the scope of this paper.

Although the molecular gas probed by the absorption only contains a tiny fraction of the 
total molecular content of the galaxy, it may be directly responsible for fueling star formation 
in the galaxy and may hence play a crucial role in the galaxy's evolution. Several observational 
attributes bolster this claim: (i)~the absorbing gas has been heavily enriched by the host 
galaxy, (ii)~its kinematics indicate that the gas is bound to the galaxy, and (iii)~it is 
relatively cold and may hence more readily fuel star formation.

\begin{figure}[t!]
\includegraphics[width=0.5\textwidth]{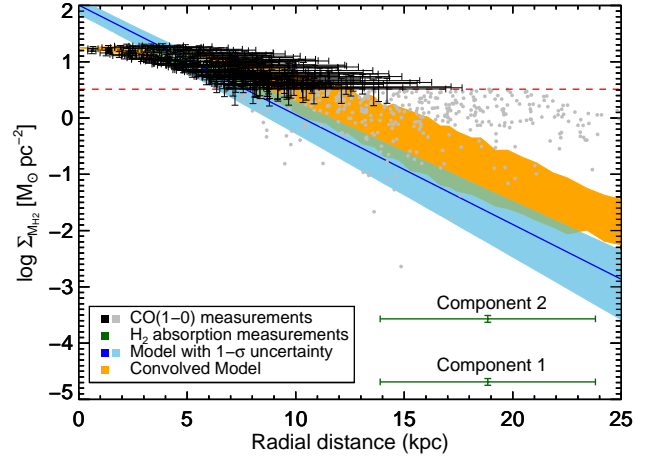}
\caption{Surface density of molecular hydrogen as a function of radius. The red dashed line
indicates the 2$\sigma$ detection limit of the ALMA data. Integrated CO(1$-$0) emission
features of $\geq$ 2-$\sigma$ significance are plotted in black with associated error
bars, whereas those below this limit are shown in gray. The blue line marks the
exponential disk model used for the MCMC analysis. This model is convolved with the ALMA 
beam (orange region), showing the excellent agreement between the model fit and the data. 
The surface densities computed from the H$_2$ absorption lines are plotted at
the distance the line of sight crosses the plane of the disk, showing that neither absorption 
component trace the same molecular phase responsible for the CO(1$-$0) 
emission.}
\label{fig:surfh2}
\end{figure}

\section{Concluding Remarks}

We have used ALMA to obtain the first detection of molecular emission from a galaxy selected to be
nearby a projected background quasar. The detected CO(1$-$0) emission yields a molecular gas mass of 
$\approx 4.2 \times 10^9 \,M_\odot$, larger than the upper limit on the atomic gas mass in this system.
The large inferred gas depletion time suggests that the molecular gas is mostly 
in a diffuse phase, rather than in dense giant molecular clouds. By combining emission and absorption studies, 
we find that the strongest absorption component does not arise in the disk of the galaxy, but instead is likely to 
stem from a diffuse CGM that is nearly impossible to detect directly through line emission or optical imaging.
Because of its metal enrichment and velocity structure, we assert this CGM gas has been enriched by the 
galaxy and could be fueling star formation in the galaxy in the future. This study highlights the synergy between
molecular emission studies and absorption studies to evaluate the cycle of gas within and surrounding galaxies.

\acknowledgements
We would like to thank H.-W. Chen for providing the B-band image of PKS 0439$-$433. M.N. and J.X.P. 
acknowledge support from NSF award AST-1109452, N.K. from the Department of Science and Technology 
through a Swarnajayanti Fellowship (DST/SJF/PSA-01/2012-13), and L.C. from YDUN grant DFF~4090-00079. 
The research leading to these results has received funding from the European Research Council under 
the European Union's Seventh Framework Program (FP7/2007-2013)/ERC Grant agreement~No.~EGGS$-$278202.

This paper makes use of the following ALMA data: ADS/JAO.ALMA\#2013.1.01178.S. ALMA is a partnership 
of ESO (representing its member states), NSF (USA) and NINS (Japan), together with NRC (Canada) and 
NSC and ASIAA (Taiwan) and KASI (Republic of Korea), in cooperation with the Republic of Chile. The 
Joint ALMA Observatory is operated by ESO, AUI/NRAO and NAOJ. The National Radio Astronomy Observatory 
is a facility of the National Science Foundation operated under cooperative agreement by Associated 
Universities, Inc.
\bibliography{bib}

\end{document}